\documentclass[11pt]{article}
\usepackage{moriond,epsfig}
\usepackage{capt-of}

\begin{document}
\newcommand{\etappee}	{\eta \to \pi^+ \pi^- e^+ e^- }
\newcommand{\etapp}	{\eta \to \pi^+ \pi^-}
\newcommand{\aphi}	{\mathcal{A}_{\phi}}
\newcommand{\pp}	{\pi^+ \pi^- }
\newcommand{\ba}{\begin{eqnarray}}
\newcommand{\ea}{\end{eqnarray}}
\newcommand{\be}{\begin{equation}}
\newcommand{\ee}{\end{equation}}
\newcommand{\bnona}{\begin{eqnarray*}}
\newcommand{\enona}{\end{eqnarray*}}
\newcommand{\non}{\nonumber}
\newcommand{\D}{\displaystyle}
\newcommand{\dd}{{\rm d}}
\newcommand{\MeV}{\mathrm{\ MeV}}
\newcommand{\GeV}{\mathrm{\ GeV}}
\newcommand{\BR}{\mathrm{BR}}

\vspace*{4cm}
\title{KLOE RESULTS ON LIGHT MESON PROPERTIES}

\author{FEDERICO NGUYEN for the KLOE COLLABORATION~\footnote{
\scriptsize
F.~Ambrosino, A.~Antonelli, M.~Antonelli, F.~Archilli, C.~Bacci,
P.~Beltrame, G.~Bencivenni, S.~Bertolucci, C.~Bini, C.~Bloise,
S.~Bocchetta, F.~Bossi, P.~Branchini, P.~Campana, G.~Capon,
T.~Capussela, F.~Ceradini, S.~Chi, G.~Chiefari, P.~Ciambrone,
F.~Crucianelli, E.~De~Lucia, A.~De~Santis, P.~De~Simone,
G.~De~Zorzi, A.~Denig, A.~Di~Domenico, C.~Di~Donato,
B.~Di~Micco, A.~Doria, M.~Dreucci, G.~Felici, A.~Ferrari,
M.~L.~Ferrer, S.~Fiore, C.~Forti, P.~Franzini, C.~Gatti,
P.~Gauzzi, S.~Giovannella, E.~Gorini, E.~Graziani, W.~Kluge,
V.~Kulikov, F.~Lacava, G.~Lanfranchi, J.~Lee-Franzini, D.~Leone,
M.~Martini, P.~Massarotti, W.~Mei, S.~Meola, S.~Miscetti, M.~Moulson,
S.~M\"uller, F.~Murtas, M.~Napolitano, F.~Nguyen, M.~Palutan, E.~Pasqualucci,
A.~Passeri, V.~Patera, F.~Perfetto, M.~Primavera, P.~Santangelo,
G.~Saracino, B.~Sciascia, A.~Sciubba, A.~Sibidanov, T.~Spadaro,
M.~Testa, L.~Tortora, P.~Valente, G.~Venanzoni, R.Versaci, G.~Xu.
}}

\address{INFN ``Roma TRE'', Roma (Italy) -- \emph{e-mail address: nguyen@fis.uniroma3.it}}

\maketitle\abstracts{
The KLOE experiment operating at the $\phi$--factory DA$\Phi$NE
has collected an integrated luminosity of about 2.5 fb$^{-1}$ and
250 pb$^{-1}$, on and off the $\phi$ meson peak respectively.
Recent results achieved from studying properties of pseudoscalar and scalar mesons
are presented.
}

\section{Introduction: KLOE detector}
\label{sec:intro}
The KLOE detector consists of
a cylindrical drift chamber~\cite{Adinolfi:2002uk}
(3.3~m length and 2~m radius),
surrounded by a calorimeter~\cite{Adinolfi:2002zx} made of lead and
scintillating fibers.
The detector is inserted in a superconducting coil
producing a magnetic field $B$=0.52~T.
Large angle tracks from the origin ($\theta>45^\circ$) are
reconstructed with momentum resolution $\sigma_p/p=0.4\%$.
Photon energies and times are measured with
resolutions of $\sigma_E/E=5.7\%/\sqrt{E({\rm GeV})}$ and
$\sigma_t=57\mbox{ ps}/\sqrt{E({\rm GeV})}\oplus 100$ ps.

\section{Pseudoscalar Mesons}
KLOE collected a statistics of about $10^8~\eta$ and $5\times10^5~\eta^\prime$
events, produced through magnetic dipole transitions, $\phi\to\eta\gamma$
and $\phi\to\eta^\prime\gamma$, with full reconstruction of decay products.
\paragraph{\boldmath{$\eta\to\pi^+\pi^- e^+e^-$}.}
CP violating (CPV) mechanisms can be tested
in the decays of the pseudoscalar mesons
by measuring an asymmetry $\aphi$ in the angle $\phi$ between the $e^+e^-$ and
$\pi^+\pi^-$ planes in the meson rest frame.
In the $\eta$ case, a nonzero $\aphi$ value would signal
CPV dynamics not directly related to most widely studied
flavour changing neutral current processes.
Furthermore, possible contributions~\cite{Geng:2002ua,Gao:2002gq}
beyond the Standard Model 
can raise $\aphi$ up to $\mathcal{O}(10^{-2})$.

The $\etappee$ analysis~\cite{Ambrosino:2008cp} is based on a
sample of 1.7 fb$^{-1}$. The event selection
consists of the requirement of one photon of $E > 250$ MeV energy,
namely the monochromatic photon from $\phi\to\eta\gamma$, and
four charged tracks coming from the interaction region.
Mass assignment for each track is done
using time of flight of the charged particles measured in the calorimeter.
Background sources are due to \emph{$\phi$ decays}:
$\phi\to\pi^+\pi^-\pi^0$ or $\eta\to\pi^+\pi^-\pi^0$ processes
with $\pi^0$ Dalitz decay; or \emph{continuum processes}:
$e^+e^-\to e^+e^-\gamma$ events with $\gamma$
conversions, 
studied using data taken at $\sqrt{s}=1$ GeV. 
The contamination is evaluated by fitting the sidebands of the
$M_{\pi\pi ee}$ data spectrum with background components
after loose cuts on the kinematic fit $\chi^2$ and on the sum of
momenta of the charged particles.
Signal events are computed after rejecting $\gamma$ conversions,
and from the fit the branching ratio
is evaluated: BR$(\etappee)=(26.8\pm0.9_{stat}\pm0.7_{sys})\times10^{-5}$.
\begin{figure}[htbp]
\begin{center}
\mbox{
\epsfig{figure=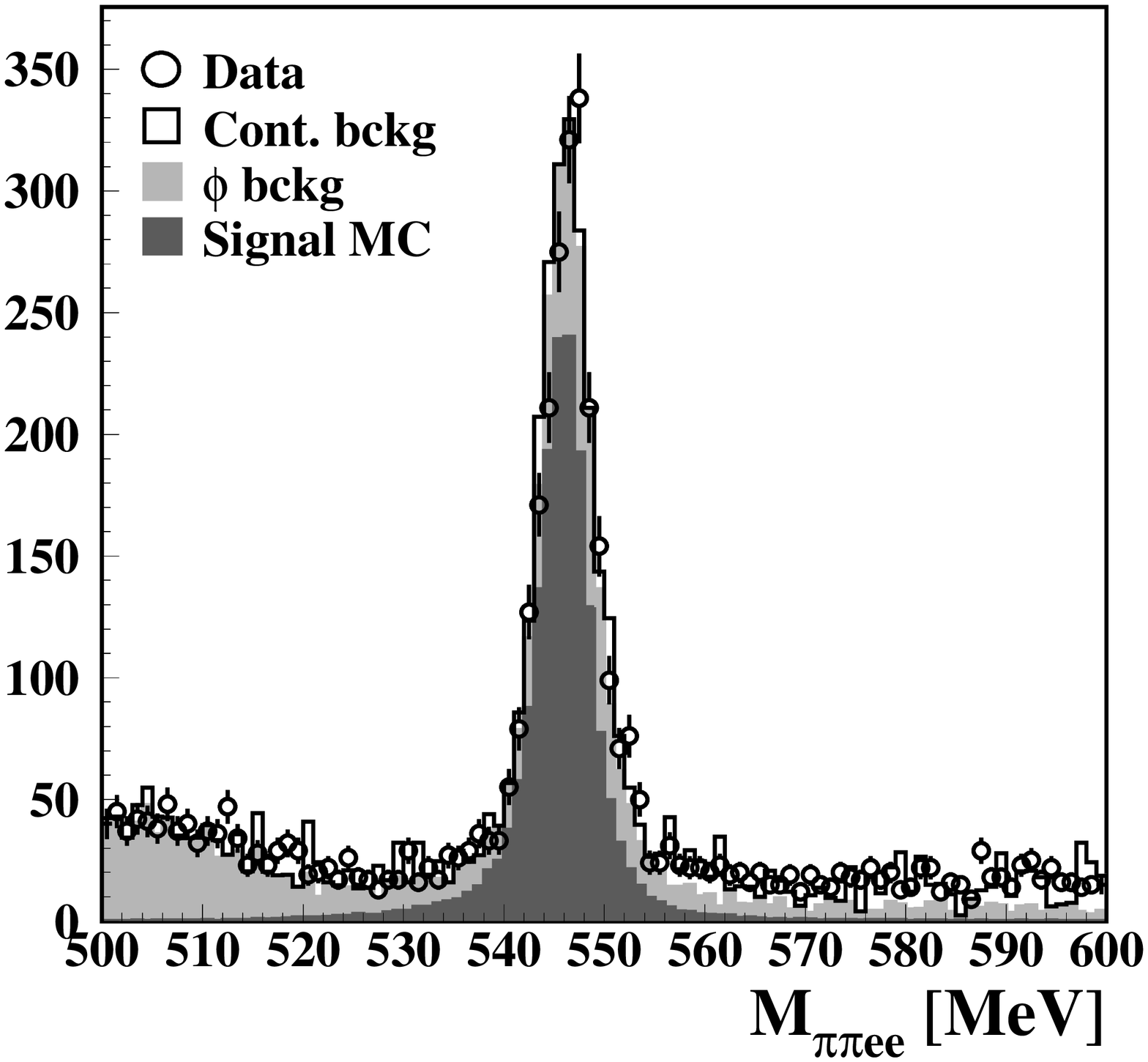,width=5cm}
\hspace*{\fill}
\epsfig{figure=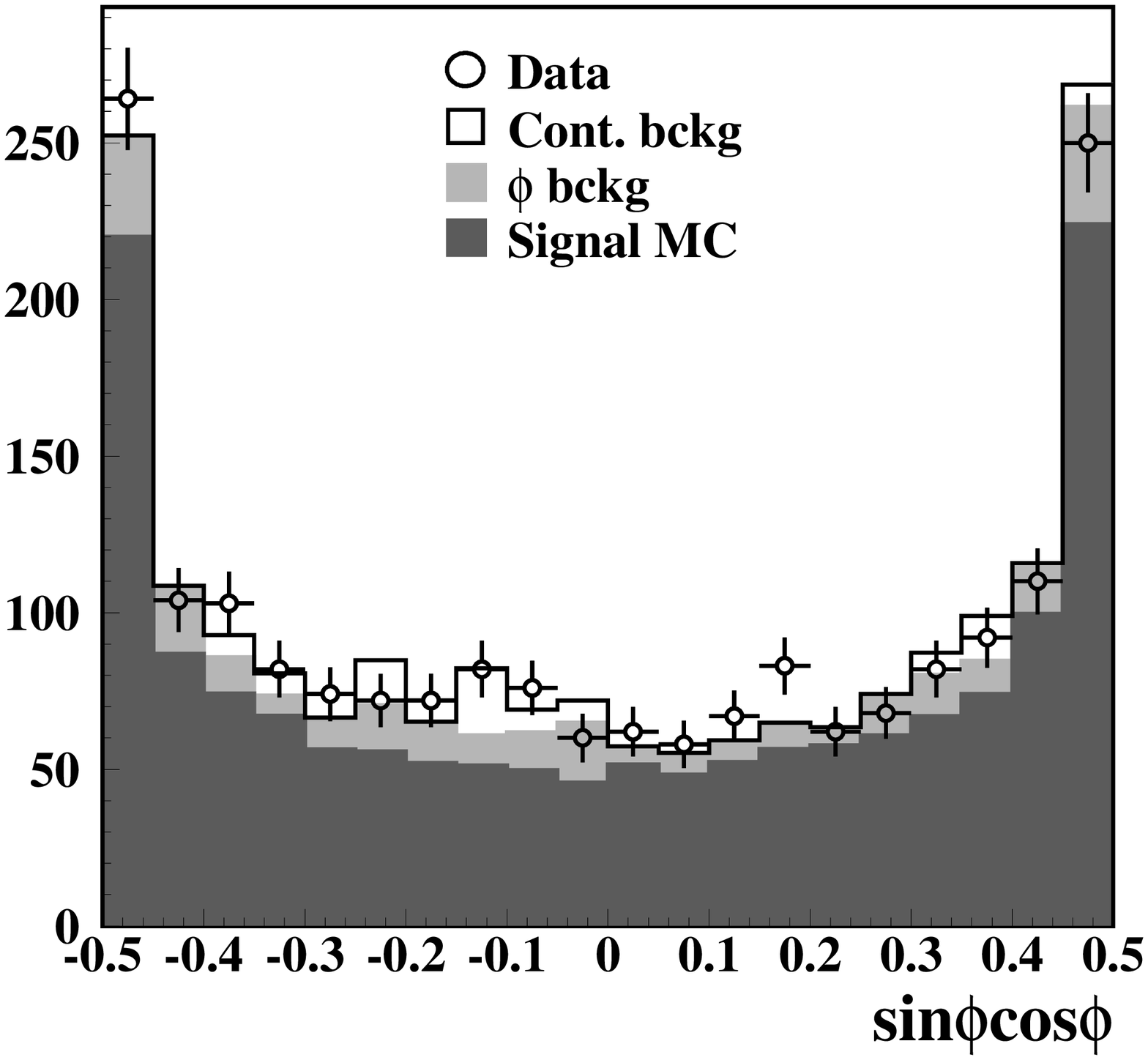,width=5cm}
\hspace*{\fill}
\epsfig{figure=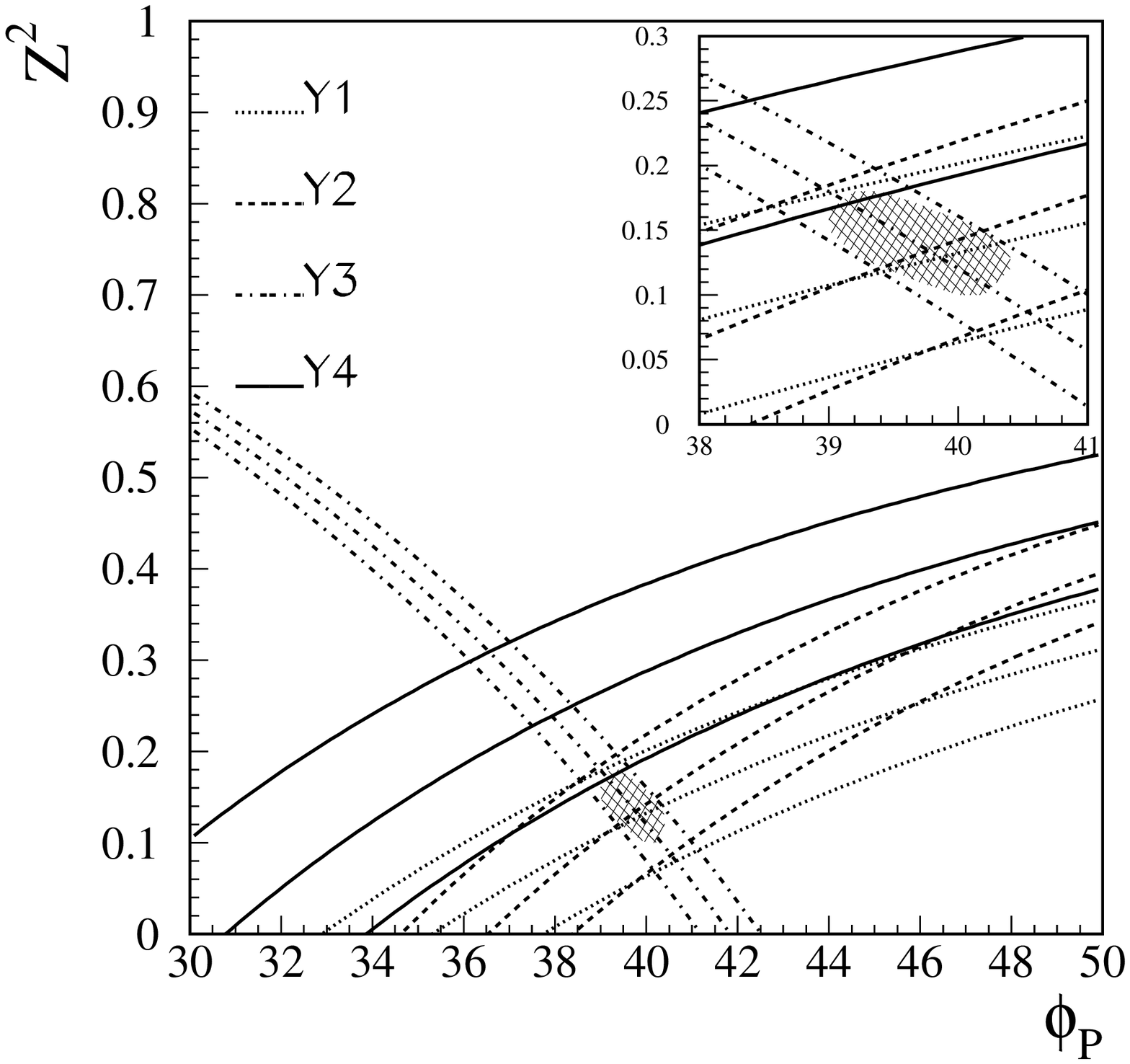,width=5.9cm}
}
\caption{\label{fig:1}\footnotesize{Left: $M_{\pi\pi e e}$ distributions
around the $\eta$ mass resulting from the fit: $\chi^2/dof=32.5/30$.
Middle: $\sin\phi\,\cos\phi$ distributions for $535 < M_{\pi\pi e e}<555$ MeV.
Right: $Z_{\eta^\prime}^2$ vs. $\varphi_P$,
with constraints from $Y1=\Gamma(\eta^\prime\to\gamma\gamma)/
\Gamma(\pi^0\to\gamma\gamma)$, $Y2=\Gamma(\eta^\prime\to\rho\gamma)/
\Gamma(\omega\to\pi^0\gamma)$, $Y3=R_\phi$ and $Y4=\Gamma(\eta^\prime\to\omega\gamma)/
\Gamma(\omega\to\pi^0\gamma)$ measurements.}}
\end{center}
\end{figure}
The asymmetry $\aphi$ is measured from momenta of the charged particles and
background subtracted. From the selected sample of 1555 events, the first
measurement of $\aphi$ performed is found compatible with zero at the 3\% level:
$\aphi = (-0.6\pm2.5_{stat}\pm1.8_{sys})\times10^{-2}$. Left and middle panels of
Fig.~\ref{fig:1} 
show the $M_{\pi\pi ee}$ spectrum after the fit and 
the $\sin\phi\,\cos\phi$ distribution in the signal region.

\paragraph{\boldmath{$\eta$-$\eta^\prime$} mixing.}
From the selection of $\phi\to\eta^\prime\gamma\to\pi^+\pi^-7\gamma$
and $\phi\to\eta\gamma\to7\gamma$ events out of a sample of
$430\mbox{ pb}^{-1}$, the ratio $R_\phi\equiv
\Gamma(\phi\to\eta^\prime\gamma)/\Gamma(\phi\to\eta\gamma)$
is measured. $\eta$, $\eta^\prime$
are known to be linear combinations of strange, $s\bar{s}$,
and nonstrange, $n\bar{n}$, quarks. Moreover the $\eta^\prime$
could contain a gluonium component $GG$, so that\\
\parbox{0.45\textwidth}{\begin{eqnarray*}
&|\eta^\prime\rangle& =  X_{\eta^\prime}|n\bar{n}\rangle
+Y_{\eta^\prime}|s\bar{s}\rangle+Z_{\eta^\prime}|GG\rangle\\[1.5ex]
&|\eta\rangle& =  \cos\varphi_P|n\bar{n}\rangle
-\sin\varphi_P|s\bar{s}\rangle
\end{eqnarray*}}\kern1.3cm
\parbox{0.35\textwidth}{\begin{eqnarray*}
X_{\eta^\prime} &=& \cos\varphi_G\sin\varphi_P\\
Y_{\eta^\prime} &=& \cos\varphi_G\cos\varphi_P\\
Z_{\eta^\prime} &=& \sin\varphi_G
\end{eqnarray*}}\\
where $\varphi_P$ is the pseudoscalar
mixing angle.
A combined fit~\cite{Ambrosino:2006gk}
of the KLOE measured value of $R_\phi\propto\cot^2\varphi_P\cos^2\varphi_G$
with available measurements of $\Gamma(\eta^\prime\to\rho\gamma)
/\Gamma(\omega\to\pi^0\gamma)$, $\Gamma(\eta^\prime\to\gamma\gamma)
/\Gamma(\pi^0\to\gamma\gamma)$ and $\Gamma(\eta^\prime\to\omega\gamma)/
\Gamma(\omega\to\pi^0\gamma)$, 
yields the following results, if $Z_{\eta^\prime}^2$ is fixed or not:
\begin{eqnarray}
\label{eq:3}
\varphi_P ~=~ (41.5^{+0.6}_{-0.7})^\circ, & Z_{\eta^\prime}^2=0\mbox{ fixed}, &~~ P_{\chi^2} = 0.01,\\
\varphi_P ~=~ (39.7\pm0.7)^\circ, & Z_{\eta^\prime}^2=0.14\pm0.04, &~~ P_{\chi^2} = 0.49.
\label{eq:4}
\end{eqnarray}
A $3\sigma$ evidence is found for gluonium content in $\eta^\prime$, as shown in Fig.~\ref{fig:1} right panel.
Further analyses, including the new measurement~\cite{Ambrosino:2008gb}
of the $\BR(\omega\to\pi^0\gamma)$
done by KLOE, confirm the gluonium content
in the $\eta^\prime$ wavefunction. 

\section{Scalar Mesons}
The still unresolved structure of these states is studied either through electric dipole transitions
such as $\phi\to a_0(980)\gamma$ and looking at the mass spectrum 
of the scalar meson decay products, or with the search for processes like
$\phi\to [a_0(980)+f_0(980)]\gamma\to K \bar{K}\gamma$ and $\gamma\gamma\to\sigma(600)\to\pi\pi$.

\paragraph{\boldmath{$\phi\to a_0(980)\gamma\to\eta\pi^0\gamma$}.}
Two independent analyses~\cite{Ambrosino:2009py}
using $\eta\to2\gamma$ or $\eta\to\pi^+\pi^-\pi^0$ decays
are performed from a sample of $410\mbox{ pb}^{-1}$.
Both analyses share the requirement of five photons from the interaction point.
The selection of also two tracks of opposite charge, while less
efficient for $\eta\to\pi^+\pi^-\pi^0$ events, has
a selected sample with smaller background
than from the $\eta\to2\gamma$ channel.
The absence of
a major source of interfering background allows to obtain the branching
fraction directly from event counting. Table~\ref{tab:1} shows that
the two samples lead to consistent branching ratio values, thus a combined
fit of the two spectra is performed. The couplings, fitted according
to the Kaon Loop~\cite{Achasov:1997ih} and the No Structure~\cite{Isidori:2006we}
models, point to a 
total width in the range $[80\div105]$ MeV and to a sizeable $s\bar{s}$ content
of the $a_0(980)$.
\begin{table}[htbp]
  \begin{minipage}{0.4\linewidth}
    \centering
    \renewcommand{\arraystretch}{1.5}
		 {\scriptsize
		   \begin{tabular}{@{}lcc@{}}
		     \hline
		     channel features    &  $\eta\to\gamma\gamma$ & $\eta\to\pi^+\pi^-\pi^0$ \\
		     \hline
		     signal efficiency & $40\%$ & $20\%$ \\
		     $B/(S+B)$  & $50\%$ & $14\%$ \\
		     $\BR(\phi\to\eta\pi^0\gamma)\times10^5$ & 
		     $7.01(10)(20)$ &
		     $7.12(13)(22)$\\
		     \hline
		     \hline
		     fit parameter & Kaon Loop & No Structure \\
		     \hline
		     $m_{a_0}$ (MeV) & $982.5(1.6)(1.1)$ & $982.5$ (fixed) \\
		     $g_{a_0K^+K^-}$ (GeV) & $2.15(6)(6)$ & $2.01(7)(28)$ \\
		     $g_{a_0\eta\pi}$ (GeV) & $2.82(3)(4)$ & $2.46(8)(11)$ \\
		     $g_{\phi a_0\gamma}$ (GeV$^{-1}$) & -- & $1.83(3)(8)$ \\
		     $\chi^2/dof$ & $157.1/136$ & $140.6/133$ \\
		     CL & 10.4\% & 30.9\% \\
		   \end{tabular}
		 }
		 \caption{\footnotesize Results from the two independent analyses (first three rows) and
		 from the combined fit of the two spectra (second six rows).}
		 \label{tab:1}
  \end{minipage}
  \hspace*{1.5cm}
  \begin{minipage}{0.45\linewidth}
    \centering
    \vspace*{-0.4mm}
    \epsfig{figure=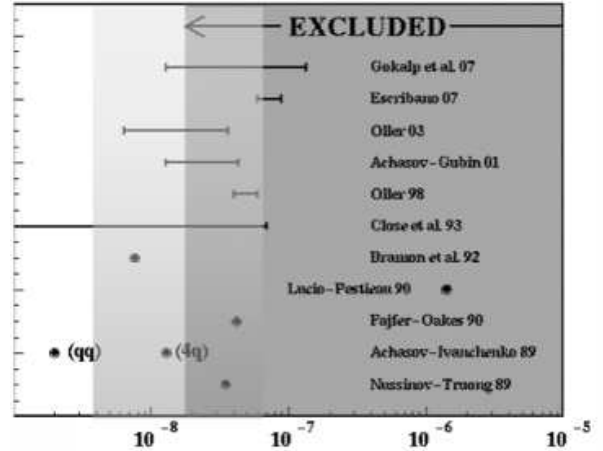,width=1.1\linewidth}
    \vspace*{-8mm}
    \captionof{figure}{\footnotesize $\BR(\phi\to K\bar{K}\gamma)$:
      excluded region (dark band) at 90\% CL compared with theoretical estimates
      and the KLOE prediction (light band).}
    \label{fig:2}
  \end{minipage}
\end{table}

\paragraph{search for \boldmath{$[a_0(980)+f_0(980)]\to K\bar{K}$}.}
This process has never been observed. Detection of
a $K_SK_S$ pair is a clear signature of a $K\bar{K}$ pair with quantum numbers $J^{PC}=0^{++}$,
thus the $\phi\to K_S K_S\gamma\to 2(\pi^+\pi^-)\gamma$ process~\cite{Ambrosino:2009rg}
is searched for,
where the main background are the resonant
$e^+e^-\to\phi\gamma\to K_S K_L\gamma$ and the continuum $e^+e^-\to\pi^+\pi^-\pi^+\pi^-\gamma$
processes.
From an integrated luminosity of 2.2 fb$^{-1}$, 5 candidate events are found in data,
whereas 3 events are expected from Monte Carlo background samples.
This leads to $\BR(\phi\to K\bar{K}\gamma)<1.9\times10^{-8}$ at the 90\% CL.
Fig.~\ref{fig:2} shows the predictions of various theoretical models for
the branching ratio. Some of them are
excluded by this measurement.
The present upper limit is consistent with the $\BR(\phi\to K\bar{K}\gamma)$ prediction
computed with $a_0(980)$~\cite{Ambrosino:2009py},
$f_0(980)$~\cite{Ambrosino:2005wk,Ambrosino:2006hb} couplings measured by KLOE.

\paragraph{search for \boldmath{$\gamma\gamma\to\sigma(600)\to\pi^0\pi^0$}.}
While there is a long debate on the observation~\cite{Aitala:2000xu,Ablikim:2004qn}
of the $\sigma(600)$ as a bound $\pp$ state, 
there is no direct evidence for the $\sigma(600)\to\pi^0\pi^0$ decay.
At DA$\Phi$NE, the detection of the process $e^+e^-\to e^+e^-\pi^0\pi^0$
implies~\cite{npipo} the intermediate process $\gamma\gamma$ to a scalar meson state.
From a sample of 11 pb$^{-1}$ of data
taken at $\sqrt{s}=1$ GeV, the feasibilty study of the
$\pi^0\pi^0\to4\gamma$ invariant mass ($M_{4\gamma}$) spectrum -- without tagging $e^+$ or $e^-$ in the
final state -- is performed.
Preliminary results show an excess of events in the $M_{4\gamma}$
region below 400 MeV, difficult to be explained by 
$\phi$ decays or $e^+e^-$ annihilation processes. This preliminary work is
encouraging and motivates the analysis extension
to the whole sample of 240 pb$^{-1}$ collected
by KLOE at $\sqrt{s}=1$ GeV.

\section{Conclusions}
The KLOE data set, together with a precise simulation
of the detector response and of a large number of several processes -- $\phi$ decays and
continuum $e^+e^-$ annihilations --
allowed for an extensive study
of properties of scalar and pseudoscalar mesons:
\begin{itemize}
\item the most precise $\BR(\etappee)$ determination and the first
measurement of the asymmetry between $\pp$ and $e^+e^-$ planes, found
compatible with zero;
\item evidence of valence gluons in the $\eta^\prime$ wavefunction
to three standard deviations;
\item good agreement between two $\eta\pi^0\gamma$ analyses with
different systematics, where the combined fit of the two spectra
points to a sizeable strange quark content in the $a_0(980)$;
\item upper limit of the $\BR(\phi\to K\bar{K}\gamma)$ at 90\% CL
with the whole KLOE statistics;
\item excess of events with respect to known backgrounds in the $4\gamma$ invariant mass.
\end{itemize}
The programme of the KLOE-2 experiment, expected to roll in the $e^+e^-$ interaction
region of DA$\Phi$NE in the second half of 2009,
includes meson spectroscopy with $\gamma\gamma$
interactions, improved measurements in the $\eta$-$\eta^\prime$ sector and
observation of the $K\bar{K}\gamma$ final states.

\end{document}